# Implementing result-based agri-environmental payments by means of modelling

Bartosz Bartkowski[1], Nils Droste[2], Mareike Ließ[3], William Sidemo-Holm[4], Ulrich Weller[3], Mark V. Brady[4,5]

[1] UFZ – Helmholtz Centre for Environmental Research, Department of Economics, Leipzig, Germany

[2] Department of Political Science, Lund University, Sweden

[3] UFZ – Helmholtz Centre for Environmental Research, Department of Soil System Science, Leipzig, Germany

[4] Center for Environmental and Climate Research, Lund University, Sweden

[5] Swedish University of Agricultural Sciences, Lund, Sweden

**Abstract**

From a theoretical point of view, result-based agri-environmental payments are clearly preferable to action-based payments. However, they suffer from two major practical disadvantages: costs of measuring the results and payment uncertainty for the participating farmers. In this paper, we propose an alternative design to overcome these two disadvantages by means of modelling (instead of measuring) the results. We describe the concept of model-informed result-based agri-environmental payments (MIRBAP), including a hypothetical example of payments for the protection and enhancement of soil functions. We offer a comprehensive discussion of the relative advantages and disadvantages of MIRBAP, showing that it not only unites most of the advantages of result-based and action-based schemes, but also adds two new advantages: the potential to address trade-offs among multiple policy objectives and management for long-term environmental effects. We argue that MIRBAP would be a valuable addition to the agri-environmental policy toolbox and a reflection of recent advancements in agri-environmental modelling.

# 1 Introduction

Agricultural production is entangled in several challenges that require appropriate design of institutional responses. Agriculture depends on an intact, functioning environment; but it also causes environmental damage. On the one hand, farmers are heavily affected by environmental change, e.g. climate change (Challinor et al., 2014; Peichl et al., 2019) and land degradation (Nkonya et al., 2016). On the other hand, agricultural production is a major source of multiple environmental pressures, including greenhouse gas emissions, soil erosion, ground and surface water pollution and landscape degradation (Campbell et al., 2017). Despite the recognition that agriculture is generating serious environmental problems, they persevere and in many cases are worsening (Springmann et al., 2018). This is often due to inadequate specification of property rights and the associated asymmetrical, inefficient and potentially unjust distribution of costs and benefits among farmers and other members of society (Bartkowski et al., 2018). From an economic point of view, agricultural produce and associated profits are private benefits, whereas the environmentally harmful impacts of agriculture are externalities (public bads to society and other stakeholders) that are not borne by farmers and hence in need of internalization in farmers' decisions. This includes the need to incentivize the provision of public goods (positive externalities such as carbon storage in soils), which risk being underprovided if there is no compensation to the farmer.

The most common policy instrument in this context is agri-environmental payments, a form of payments for ecosystem services (PES) (Engel, 2016): farmers voluntarily enter contracts under which they agree to change their management in a way that is assumed to benefit the environment. In exchange, they receive pre-defined payments. There are two general variants of agri-environmental payments – action-based and result-based schemes.[1] *Action-based schemes,* which are much more widespread today, offer farmers a uniform payment within a specified area or region such as a watershed for adopting specific management practices or environmentally beneficial actions. *Result-based schemes*, on the other hand, offer payments conditional on achieving a result, i.e. a quantifiable environmental objective, while the choice of actions to achieve the result can be up to the participating farmers. The defining characteristic of a result-based scheme is that the payment is based on a quantified result, and therefore implies the possibility of farmers receiving different payments for the same actions.

---

[1] Note that different terms are used in the literature: action-based can be referred to as input- and measure-based or action-oriented payments/schemes, while result-based are referred to as performance-, outcome-, output-, success-based or -oriented payments/schemes, or objective-driven or payment-by-result schemes.

Most agri-environment and climate measures within the European Union's (EU) Common Agricultural Policy (CAP) framework are action-based payment schemes (Burton and Schwarz, 2013). A growing body of literature indicates that these schemes perform poorly; while the CAP's agri-environmental programmes have been shown to slightly improve the state of European agroecosystems (e.g. Batáry et al., 2015), action-based schemes lack the important sensitivity to local conditions (Kleijn et al., 2011) and they often fail in providing the ecological benefits they are supposed to bring about (Burton and Schwarz, 2013; Dupraz and Guyomard, 2019). Moreover, they are often cost-ineffective (Wätzold et al., 2016). Overall, the literature indicates that the lack of evidence-based links between the implementation of particular practices on particular farms is the root of the poor performance of action-based schemes.

A large body of literature exists on the relative strengths and weaknesses of action-based and result-based agri-environmental payments (Börner et al., 2017; Burton and Schwarz, 2013; Engel, 2016; Engel et al., 2008). This literature shows that result-based payments are clearly preferable from a theoretical point of view because they provide incentives to farmers to enrol the most suitable land, thus ascertaining goal attainment and preventing adverse selection; they have low informational requirements for the regulator (which, however, goes along with potentially high information rents for the farmers (White and Hanley, 2016)); they are cost-effective and dynamically efficient by providing incentives to innovate and drive down the costs of goal attainment over time. Furthermore, it has been pointed out that by being less prescriptive and by rewarding inventiveness they may increase farmer engagement and lead to an internalization of the scheme's goals by the farmers (Burton and Schwarz, 2013). Why then are not result-based schemes more prevalent?

Result-based payments score significantly worse than action-based payments in terms of practicability, which explains their low prevalence in practice: first, they require, ostensibly, sophisticated monitoring and measurement of results (Zabel and Roe, 2009); second, they are less attractive to farmers than action-based payments due to the associated uncertainty of payment, as "an individual's performance also depends on external environmental effects such as weather influences" (Zabel and Roe, 2009, p. 131; see also Drechsler, 2017; Derissen and Quaas, 2013). Furthermore, conventional result-based payments provide incentives to enrol land where the required effect is already fulfilled or close to fulfilment (Uthes and Matzdorf, 2013), as usually the payments are not based on a change compared to the status quo, but rather on the absolute level of achievement.² Nonetheless, result-based agri-environmental payments are widely considered the way forward in Europe (e.g. Cullen et al., 2018; Mann, 2018). For instance, the European Commission has issued a handbook on designing and implementing result-based schemes within the CAP (Keenleyside et al., 2014).

The main obstacle posed by result-based schemes compared to action-based schemes is therefore practical: the measurement of the results of farmers' chosen measures. In the context of nonpoint-source pollution, Sidemo-Holm et al. (2018) have demonstrated that the practical shortcomings of result-based agri-environmental payment schemes can be alleviated by using models instead of direct measurement to determine the farmer's achieved result (see also Talberth et al., 2015). In this paper, we aim to develop further the idea of using modelling for solving the measurement problem and to demonstrate how uncertainties can be reduced on both sides of the transaction. To do this, we propose the design of a model-informed, result-based agri-environmental payment (MIRBAP) scheme and discuss how it would combine "the best of both worlds" of action-based and result-based payments. Furthermore, we provide a framework of how modelling and smart infrastructure can be combined for possible applications in the context of sustainable land management, and demonstrate its applicability with a concrete example in the area of soil function modelling and soil management.

The structure of the paper is as follows: in section 2 we elaborate on the relation of measurement and modelling in the context of result-based schemes. In section 3 we outline the general idea of a model-informed result-based payment scheme and illustrate it with a hypothetical application in the context of soil functions. In section 4 we offer a comprehensive discussion of the relative advantages and disadvantages of our design proposal. In section 5 we conclude and suggest some areas for future research.

## 2 Measuring versus modelling results

Typically, result-based schemes are taken to mean schemes that are based on actual measurement of environmental results through monitoring. Consequently, in situations where it is infeasible to measure results directly due to either the lack of a perfect object of measure (e.g. P concentration in water) or high cost of measurements at the individual farm or field level (e.g. N concentrations in tile drains), it is also assumed that a result-based scheme is infeasible.

---

² According to Herzon et al. (2018), most result-based schemes provide payments on the basis of "the opportunity costs of the management that is considered most likely to be required to achieve the results" (p. 350). In fact, this is considered a legal requirement according to the WTO "Green Box" rules (Hasund and Johansson, 2016) and in line with EU's own Rural Development Regulation, Article 28 (Colombo and Rocamora-Montiel, 2018).

However, if a suitable proxy for measuring the environmental result is available, then this can solve the measurement problem, and open the way for broader application of result-based schemes.

So far, result-based agri-environmental schemes in the EU have adopted the first strategy for overcoming the measurement problem, by remunerating farmers based on an indicator as a proxy for the environmental result, usually biodiversity (see Table A1). Since biodiversity is usually defined by a complexity of factors, it has proven impracticable (i.e. it is infeasible or too costly) to measure explicitly the impact of farming practices on biodiversity per se. Instead, farmers' payments for biodiversity conservation are based on indicators of results rather than farmers' specific actions. The indicators include populations of particular species whose presence and abundance are known to correlate with a wider range of taxa. A conclusion that can be drawn from the EU inventory of result-based schemes is that these schemes are not dependent on measurement per se, but on indicators or proxies that have predictable relationships with the environmental objective. So far, result-based schemes in Europe are all targeting biodiversity – while other environmental public goods such as soil functions and regulating ecosystem services (e.g. water quality or flow regulation) are only addressed by means of action-based schemes (if at all).

In cases where the feasibility or cost of measurement is the barrier to result-based schemes, modelling offers a solution. In a model-informed result-based payment scheme as described in detail in the following section, environmental results are predicted rather than measured. Models can synthesize knowledge about agroecosystems and process that knowledge to predict environmental benefits that are not feasible to measure, e.g. the influence of agricultural management on the dynamics of soil functions (Vogel et al., 2018) and nonpoint-source pollution (Strauch et al., 2013). In order to do so, the model needs to describe the particular system's state and dynamics and its alteration by external factors (e.g. agricultural practices and climate), and thus provide predictions of the environmental effects of management actions.

Models can of course never perfectly represent a complex system such as the interface between agriculture and the environment. There are always things going on in such a system that cannot be modelled because of data and resource shortages, and inherent stochasticity. How well a model resembles reality is revealed by the uncertainty of its predictions. It is therefore desirable that model predictions have as low uncertainty as possible when used for policy purposes, such as result-based payment schemes. In contrast, measured results are ideally true values. The only uncertainty lies in the accuracy of the measurement methods – which may be large depending on the availability of measurement theory and technology. However, as discussed above, existing result-based payments are not based on measured results, but rather on indicators. Indicators imply high levels of uncertainty, just as modelled predictions, with respect to the reliability and stability of the relationship between the indicator and the desired result. Compared to measured indicators, models have the advantage that they can be used to estimate the uncertainty of the predictions.

To predict environmental benefits with the lowest possible uncertainty, models need to include all relevant processes, functional relationships and dependencies. Otherwise, their predictive power will be limited (Evans et al., 2013). Model development requires expert knowledge. Data is required for model building, calibration, and evaluation. Still, measuring environmental benefits directly requires far more data in comparison. Expert knowledge is required to select reasonable indicators and to interpret their meaning.

Before estimating results with models, it has to be decided whether the predictions for the eligible management practices are made with acceptably low uncertainty. A threshold for acceptable uncertainty can be decided upon initially, and as long as that threshold is exceeded, the model needs to be revised until the uncertainty is at an acceptable level, or otherwise abandoned. This process is similar to when evaluating the measurability of indicators in conventional result-based payments schemes. Instead of appraising the capacity of farmers, controllers and measurement tools, the modelling accuracy and precision are considered.

## 3 Model-informed result-based agri-environmental payments

From a theoretical point of view, result-based schemes are superior to action-based schemes (see section 1). However, from a practical perspective, result-based schemes suffer from two main shortcomings, both of which are broadly related to measurement: first, there is the general challenge to measure at acceptable cost the results whose realization is to be remunerated; and second, the result-based payment constitutes a source of uncertainty for participating farmers. The gist of the present article's argument is that both shortcomings can be overcome by means of modelling, though at a certain cost in terms of outcome uncertainty. In this section, we lay out a design of a model-informed result-based agri-environmental payment (MIRBAP) scheme. First, we present the conceptual idea in a general sense; subsequently, we illustrate it using the example of a hypothetical MIRBAP scheme targeting soil functions.

## 3.1 Conceptual proposal

The core idea is that, instead of using ex-post indicator measurements to determine the achievement of results, a MIRBAP scheme would employ models to predict the results. These predicted environmental effects would then be the basis for determining payments. In what follows, we elaborate on how the information would flow between different inputs at different stages of a MIRBAP implementation, including modelling, farmer's choices, results, and payments. Furthermore, we employ the idea of an integrated model-software application to sketch out what an actual implementation may look like.

Firstly, models need to be fed with spatially explicit data describing agronomic, ecological and biophysical features of the land, e.g. landscape structure, field size, soil type, hydrology and crop rotation, that are needed to predict the effects of management on environmental outcomes (Fig. 1, step 1). Secondly, the models are used to predict the effects of a set of management actions, e.g. reduced tillage, mowing dates, buffer strips and linear natural elements, in terms of spatially explicit environmental outcomes, e.g. changes in the provision of soil functions or biodiversity (Fig. 1, step 2). Thirdly, the modelled environmental outcomes are used as a basis for determining payments (Fig. 1, step 3). This entails that farmers are presented with payment offers for a suite of spatially explicit options of different actions (and their intensity) from which they can choose for their farm. Fourthly, the farmers choose one or more of the arrangement of actions (practices) according to their own preferences and knowledge about their fields, e.g. soil productivity, cost structures, profitability, and other types of motivation (Fig. 1, step 4). We suggest that this step can be facilitated with a software application that provides a graphical user interface (cf. Fales et al., 2016; Sturm et al., 2018). Such a software application, either web-based or mobile, could make the range of possible management choices under a particular scheme more accessible to the user (the farmer). Given the spatially explicit data input and the management choice by the farmer, the MIRBAP application would provide the farmers with predictions about environmental outcomes and entailed payments. Fifth and lastly, the scheme should be continually validated and (if necessary) updated to improve the accuracy of predicted results and effectiveness in terms of environmental outcomes (Fig. 1, step 5). This should include accounting for changes in land use which follow farmers' participation in the MIRBAP, e.g. effects on the local hydrology from planting a hedge, and validating with environmental monitoring at a larger scale.

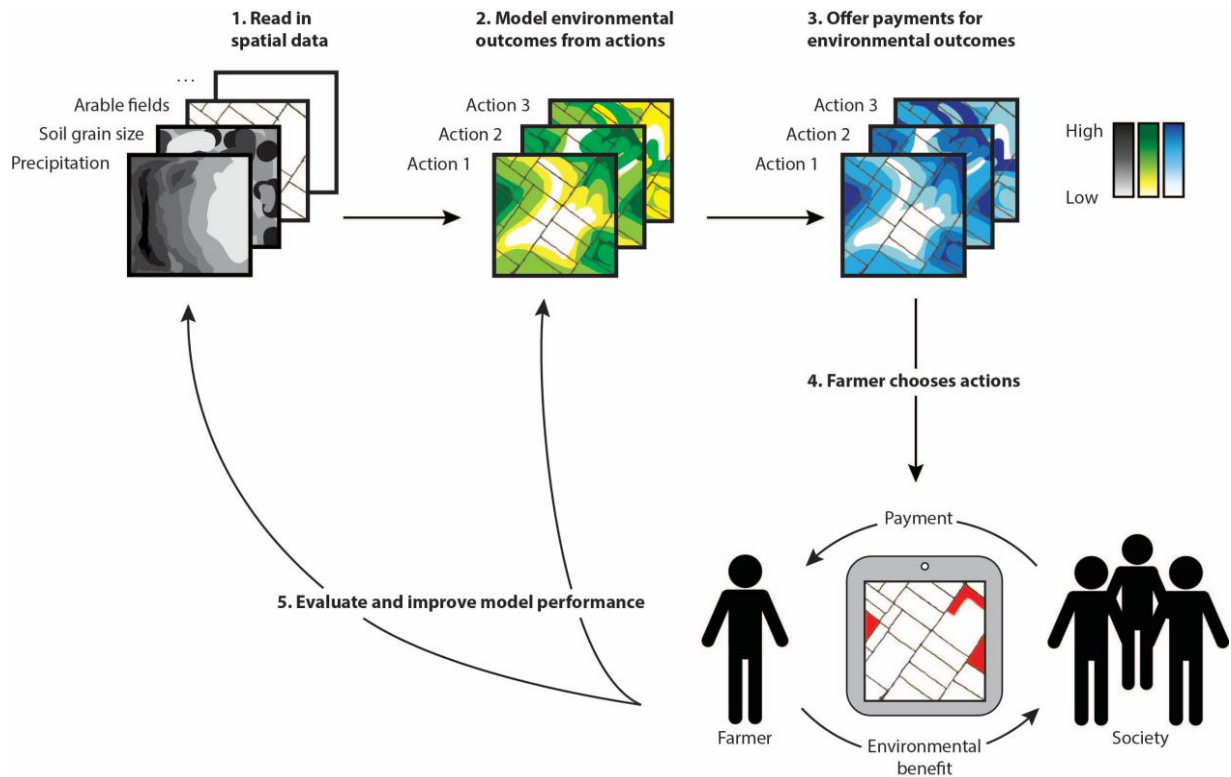

*Figure 1: Schematic outline of a model-informed result-based agri-environmental payment (MIRBAP) scheme*

From a conceptual point of view, a MIRBAP scheme constitutes a form of contract between society and participating farmers (Hagedorn, 2008). Society can benefit from the provision of (public) ecosystem services by the farmers and the agroecosystems they manage. Taking the farmers' freedom of choice into account, society (and the regulator on society's behalf) make an "offer" for the provision of specified ecosystem services (in the widest sense). Within a safe virtual space of a smart application, farmers can experiment with different options and pick an action or a set of actions that they prefer most, be it because of monetary pay-off, costs, environmental quality, or any other criteria such as intrinsic motivation. The software application would thereby become a decision-support tool to allow the farmers and regulators to make informed decisions under reduced uncertainty (for further discussion of the advantages and disadvantages, see section 4). In other words, the farmer implements the action and receives a predefined, certain payment contingent upon performing the action chosen, based on ex-ante model prediction and subject to a potential control by the regulator. This entire scheme can be re-evaluated and improved over time with regard to four main elements: quality and resolution of the input data, model prediction accuracy and precision, the set of available measures, and the corresponding payment levels.

## 3.2 MIRBAP scheme example

There cannot realistically be one model for all agri-environmental payment schemes – rather, for each specific context, a specific model is required. Currently, agri-environmental policy instruments directly addressing soils and the functions they provide are scarce in the EU (Ronchi et al., 2019; Vrebos et al., 2017). As already mentioned in section 2, virtually all existing result-based agri-environmental schemes in Europe are targeting biodiversity. Given the complexity and heterogeneity of soil systems, action-based schemes are likely to be quite ineffective in improving the provision of soil functions. At the same time, soil functions are difficult to measure on a large scale (e.g. Drobnik et al., 2018). Accordingly, MIRBAP schemes have a very large potential in the context of soils (Bartkowski et al., 2018; Vogel et al., 2018). In this section we present an example of a modelling approach that could (hypothetically) be the basis for a MIRBAP scheme in the context of soil functions. The example is based on a modelling framework currently under development within the German large-scale project BonaRes – Soil as a Resource for the Bioeconomy. We explain and discuss the steps depicted in Figure 1 in this particular context, with an emphasis on Steps 1 and 2. We do not include Steps 4 and 5, as they become relevant only in an actual application of MIRBAP in the real world.

*Step 1: Spatial data and modelling*

In order to implement and evaluate the MIRBAP application, spatial data from various sources are required that can be used to make the model-based translation of management into soil functions spatially explicit and context specific (Step 2). These essentially refer to meteorological, crop and soil data (compare Fig. 1). For the development of spatially continuous high-resolution geo-information from raw data from various sources (e.g. soil profile description, data from lab analysis, radar measurements), regionalization approaches are required that often rely on space-borne remote sensing and involve diverse modelling procedures. Models applied to generate spatially distributed rainfall and other climate data depend on the temporal and spatial scale, and include empirical statistical models as well as models of dynamic meteorology (see Srikanthan and Mcmahon, 2001 for a review). The EU's Sentinel satellite mission produces better temporal data continuity compared to previous satellite missions which may allow for the derivation of crop-specific land use classification and yield estimation, based on empirical modelling and complex data processing routines (Battude et al., 2016; Veloso et al., 2017).

Spatial soil information is often available as conventional polygon maps displaying the spatial extent of systematic units. However, as soil map development heavily relies on the individual soil surveyor's expertise, the methodology is not reproducible; spatial aggregation follows primarily optical criteria, soil types of different properties and genesis are merged. On the other hand, the state-of-the-art empirical modelling approach often relies on the same legacy soil databases (Arrouays et al., 2017), but translates expert knowledge on pedogenesis in predictor-response systems that are fitted by powerful algorithms from machine learning and spatial statistics. The approach is known as "digital soil mapping" and includes model performance evaluation and the provision of an uncertainty estimate per se. Geo raster data that approximate the soil forming factors (parent material, climate, topography, vegetation/land use) are used as predictors. They may originate from remote sensing data products and expert-based information; see Nussbaum et al. (2018) and Padarian et al. (2019) for recent applications.

*Step 2: Linking management to soil functions*

Within the MIRBAP framework, the above-mentioned geo raster data is used to feed into models linking management with environmental objectives. Within BonaRes, a process-based soil model called Bodium is currently under development. It attempts to integrate in a systemic approach (Vogel et al., 2018) the dominant processes in soils, and thus to facilitate the prediction of the multiple consequences of management practices for soil processes and, eventually, soil functions. Soil structure dynamics are an integral part of the modelled soil system. Plant roots and soil organisms build this structure, while agricultural machines alter it. The dynamic soil structure influences water and air distribution and these in turn influence plant growth and organism activity. This biotic activity is the basis for nutrient cycling and organic matter decomposition, i.e. carbon sequestration and storage. All these processes are site-specific and depend on local conditions. These conditions comprise climatic and soil factors that cannot be influenced at a medium time scale – they are so-called "inherent soil properties" (Vogel et al., 2018). This input is provided by the georeferenced data and modelling synthesized in Step 1.

To facilitate the running of Bodium, scenarios are used. Bodium translates management changes into outcomes that are evaluated according to the major soil functions: production, storage for carbon, storage and filter for water, soil biodiversity, and nutrient cycling. Under the assumption that production and nutrient cycling are essentially private goods, they would not be used as objectives for a MIRBAP scheme; however, the modelling results with respect to these two have obvious relevance for the farmer's decision to participate in the scheme and, thus, in the sense of a decision support tool. Carbon storage, water buffering for flooding, groundwater recharge, nutrient retention and biodiversity are (expressions of) soil functions that have clear societal relevance in terms of regulating and supporting ecosystem services, and modelled changes in them could therefore be used to determine agri-environmental payments. The nature of Bodium, which models multiple soil functions simultaneously, would allow for a holistic approach that better takes into account the trade-offs involved than when single functions are targeted separately. For instance, there is a well-known trade-off between reduction of herbicide use and a corresponding increase in the need for mechanical weeding and thus compaction (Böcker et al., 2019).

Within a MIRBAP scheme, the data and modelling from Steps 1 and 2 would be the basis for an online application available to the farmers, also being the source of combined spatial data describing a farmer's land eligible to participate in the agri-environmental payment scheme. Optionally, it is thinkable to allow the farmers to correct the data on the basis of standardized data input, e.g. from proximate sensors attached to the farmers' machines. These input data would then underlie the Bodium-based modelling. Changes in the soil production function would mainly inform the farmer about opportunity costs of the available actions. Changes in the other, public soil functions though would be the basis for offering payments to the farmer.

*Step 3: Payments*

For each soil function, payment rates per unit increase would be specified, according to which the farmers would receive remuneration. Generally, it is an important question for MIRBAP and any other result-based scheme how the payment rate per unit of the objective is to be determined. From a welfare economic point of view, it would be efficient to remunerate farmers for the realization of environmental benefits on the basis of the latter's marginal social value (i.e. shadow price) (Hasund, 2013). Otherwise, there is the risk of underprovision of the environmental public good in question.[3] On the other hand, however, if the payments are significantly higher than the opportunity costs (overcompensation), underprovision may result because the scheme funds are limited and more farmers could be enrolled if payments were lower (Börner et al., 2017). In the end, "the payment level determines the distribution of net gains between [ecosystem service] providers and [ecosystem service] beneficiaries" (Engel, 2016, p. 139). Of course, shadow prices of environmental public goods (ecosystem services, biodiversity etc.) can only be roughly approximated by means of economic valuation, and there is a general paucity of high-quality valuation studies (e.g. Bartkowski et al., 2015; Förster et al., 2019). This also holds in the specific context of soil valuation studies (Bartkowski et al., under review; Jónsson and Davíðsdóttir, 2016). As a first step, a MIRBAP scheme targeting soil functions may thus rather use payment levels derived from a stakeholder-based negotiation process between the farmers and representatives of the wider society or set on the basis of the opportunity costs of a typical relevant management action. Eventually, valuation estimates from specifically conducted studies could be used to inform the adaptation of payment levels.

---

[3] Also, because non-economic factors also determine the participation in agri-environmental schemes, payments alone need not necessarily guarantee socially efficient levels of participation (and, thus, provision of the public good) (Bartkowski and Bartke, 2018).

## 4 Discussion

In the last section, we presented the idea of a model-informed result-based agri-environmental payment (MIRBAP) scheme, both in a more generalized, conceptual form and in a more specific context, using an example related to soil functions. It should be noted that in the United States, a (pilot) scheme was implemented that is consistent with the MIRBAP principles as described above. It had been developed by The Nature Conservancy within a Strategic Agricultural Conservation programme in the Saginaw Bay watershed, Michigan, USA (Fales et al., 2016). As part of a large-scale conservation programme, the Great Lakes Watershed Management System (GLWMS) was developed, a model-based online decision support tool. Among other uses, the GLWMS was the basis for a pay-for-performance (i.e. result-based) pilot programme, with payments provided for various management practices on the basis of their model-estimated effects on water quality. In this section, we would like to go beyond the quantitative results and scarce discussion provided by Fales et al. (2016), and discuss the general advantages and disadvantages of MIRBAP. To guide the discussion, we use a set of policy evaluation criteria informed by the literature on agri-environmental schemes and payments for ecosystem services (of which the former are a particular instance) (Burton and Schwarz, 2013; Engel, 2016; Uthes and Matzdorf, 2013). Table 1 summarizes the comparison of MIRBAP with conventional action-based and result-based schemes (based on the authors' assessment and expertise).

*Table 1: Relative performance of action-based, result-based and MIRBAP schemes.*

*Note: This is a purely relative scale where green indicates the best relative performance with respect to a criterion, yellow is 2nd best and orange is 3rd best. If two scheme types are tied according to a criterion, there no orange coding is used (only green and yellow).*

| Criterion | Action-based | Result-based | MIRBAP |
|---|---|---|---|
| Outcome certainty | 3rd | 1st | 2nd |
| Payment certainty | 1st | 2nd | 1st |
| Additionality | 3rd | 2nd | 1st |
| Cost-effectiveness | 2nd | 1st | 1st |
| Dynamic efficiency | 3rd | 1st | 2nd |
| Farmer autonomy | 2nd | 1st | 1st |
| Multiple objectives | 2nd | 2nd | 1st |
| Long-term objectives | 2nd | 2nd | 1st |

## 4.1 Outcome certainty

From the point of view of society and the regulator (but also the intrinsically motivated farmer) the environmental effectiveness of applied measures, i.e. outcome certainty, is crucial. In this respect, MIRBAP has an obvious advantage as compared to action-based payments – while the latter are based on an implicit 'one-size-fits-all' assumption, by using spatially explicit data and modelling, MIRBAP responds to the challenges of the inherent spatial heterogeneity of agroecosystems and farming systems; it is context-specific. Also, when compared with conventional result-based payments, where the results are actually measured, modelling has advantages over measuring. First, measuring results is very costly in terms of time and resources. Instead, the same models can often be applied to large regions; e.g. ICECREAM (Rekolainen and Posch, 1993) has been used to model phosphorus leaching with field resolution from agricultural land in the whole of Sweden (Johnsson et al., 2008). Thus, models can require far less time and resources to cover large areas and many farms. Additionally, as mentioned in section 2, models can make predictions for environmental benefits that cannot be measured. Further, models make possible anticipating environmental benefits before they are generated (ex-ante evaluation). Policy makers can therefore be informed about future environmental benefits on a field level. For this, it is necessary that models forecast at an adequate temporal and spatial resolution to allow policy makers to evaluate the contribution of local actions to the probability of reaching environmental goals, and use such information to plan future commitments.

However, one must say that a MIRBAP scheme is only as good as the model(s) underlying it. The uncertainty of model predictions will vary between different environmental benefits. For example, models with good precision and accuracy have been used to predict the impact of agricultural practices on greenhouse gas emissions (Weiske et al., 2006), buffer strips in preventing nutrient emissions from leading to downstream eutrophication (White and Arnold, 2009) and the impact of various agricultural practices on water quality (Fales et al., 2016) or the provision of soil functions (see section 3.2). Other environmental benefits are more difficult to model. For example, the extinction rate of species is inherently stochastic and can hardly be modelled without a discouragingly wide range of possible outcomes (Ludwig, 1999).

Whether the modelled result-based system leads to attainment of society's environmental goals in the long-run will depend on the accuracy of the predictions. Each model is calibrated and evaluated based on measurements – therefore, modelling generally cannot be more precise than measurement. This can only be the case when the model is applied to an area for which measurement is impractical (assuming that other conditions for model application are met). Furthermore, since models required for both Step 1 and Step 2 of a MIRBAP scheme are likely to be complex and data-intensive, possible errors propagate throughout the modelling chain. They comprise uncertainties of the original input data, the models involved in spatially continuous data generation, the applied process models simulating the system dynamics, and the assessment scheme of the model outputs. Therefore, uncertainties in the obtained predictions have to be quantified. By quantifying uncertainty, models also make possible the integration of uncertainty in decision-making. Policy makers can adjust the payment so that it is higher when the uncertainty is low, and vice versa, to account for the risk that the envisaged outcome is not reached despite payments being made. In comparison, the correlative uncertainty between measured indicators and desired results in conventional result-based schemes is not quantified and thus cannot be used to improve decision making.

As a way to further deal with uncertainty, schemes can include both measured and modelled mechanisms. For instance, farmers can be paid for their modelled nutrient pollution abatement, which would be difficult to measure on a farm level. The water quality can then be measured for the entire watershed area to validate the aggregated predicted pollution abatement, and the model improved thereafter.

**4.2 Payment certainty**

One of the main advantages of action-based payment schemes over result-based payments is, from the perspective of the participating farmers, that they offer the certainty of payment – if farmers sign up for a scheme and apply the agreed-upon management actions, they will be paid. Meanwhile, in a result-based scheme, the payment is conditional on achieving the result – which, however, is not only influenced by the actions of the farmer, but also by various external factors such as extreme weather events, unexpected pest infestations, the actions of neighbours etc. Payment uncertainty is one of the areas in which MIRBAP scores better than conventional result-based payments – since payments are based on ex-ante results of modelling, payment is as certain as in the case of action-based schemes. In this sense, MIRBAP combines two main advantages of result-based and action-based schemes, respectively: while offering greater outcome certainty to society than action-based payments (though less than is the case in a conventional result-based scheme), it also provides similar payment certainty to the farmer.

**4.3 Additionality**

Additionality is a common condition in evaluations of PES schemes (Engel, 2016). In a MIRBAP scheme, the model would compare the status quo with a change. Therefore, it is unlikely that payments will be provided for results that would have been achieved (or were already achieved) in the absence of the scheme. While this could similarly be done in result-based schemes where the results are measured, by comparing the environmental outcomes before and after enrolling in a scheme, result-based schemes generally only consider results after enrolling, without much regard to how it compares to the status quo. The additionality of action-based is unknown since the results are not quantified. Using the status quo as a baseline in MIRBAP may however imply moral hazard – by creating an incentive to downgrade land before entering the scheme in order to achieve a higher increment (and thus higher payment). The alternative would be to use some general baseline, such as "good agricultural practice" for the land in question – here, the downside would be the potential inefficiency due to paying for non-additional effects (windfall gain), if the actual status quo is above the baseline. Furthermore, this approach would require the definition of a general standard to be used as a baseline – thus counteracting the context-specific nature of result-based payments. Eventually, it is an empirical question which variant is likely to be less problematic and more practicable.

**4.4 Cost effectiveness**

From a theoretical point of view, result-based payments are generally considered more cost-effective than action-based payments, especially under the assumption that information asymmetries are limited and the regulator is able to perfectly monitor the results (Melkonyan and Taylor, 2013; White and Hanley, 2016). Spatial targeting can, principally, improve cost effectiveness in both types of schemes with near-perfect information availability to the regulator (Drechsler et al., 2016; Engel, 2016). However, the regulator usually has limited knowledge of the cost structures of individual farms. Result-based payments are expected to solve this information asymmetry, because payments are only provided in exchange for the provision of environmental benefits (Burton and Schwarz, 2013). Since rational farmers will only implement measures that are profitable/utility-improving (i.e., do not generate a net loss/dis-utility) to them, they will implement measures until the Equi-marginal Principle for cost effectiveness is satisfied across all farms. Preliminary assessments of existing result-based schemes have corroborated this theoretical prediction (Matzdorf and Lorenz, 2010; White and Sadler, 2012). In contrast, action-based schemes are highly cost-ineffective, as the underlying uniform payments do not reflect spatial heterogeneity in costs and benefits, thereby making it impossible to achieve cost effectiveness (see section 4.1 on their generally questionable effectiveness). In this context, MIRBAP is largely comparable with conventional result-based schemes. The information asymmetries due to unknown costs remain, but the regulator can predict the results (outcomes) of different management actions implemented by farmers. Offered a payment per unit for the desired results, rational farmers will, as in a measured result-based scheme, be incentivized to equalize their marginal costs of implementing measures with the level of the payment, thereby guaranteeing cost effectiveness of achieving the predicted results. In this way, MIRBAP achieves spatial targeting of measures without the regulator needing to know the costs for individual farms. Rather, by serving farmers with context-sensitive, spatially explicit modelling, the translation of management choices into environmental effects will ensure the choice of cost-effective measures (because accurately modelled results will incentivize the farmer to minimize their costs in the same way as measured results do).

This relates to another issue, namely that MIRBAP is essentially a combination of action-based and result-based elements, and thus invites one to ask whether it would not be better to have an action-based scheme in which the regulator uses the model to inform farmers how much they will be paid for particular measures on their farm. There are three strong arguments speaking against this. First, individualizing payments for measures in this way contradicts the basic characteristic of an action-based scheme, that of a uniform payment for the same measure.

Second, it reduces the transparency of the scheme, in that the farmer will need to rely completely on the authorities' determination of measures and payments. Third, relatedly, it would remove the need for the farmer to understand the environmental consequences of their actions. Thus, it would maintain the feeling of being controlled by a central power rather than being empowered (Rode et al., 2015). Both would likely reduce the potential environmental benefits of a MIRBAP scheme by reducing participation and increasing administration costs in an essentially top-down designed scheme. Overall, we believe that the capacity to increase farmer engagement in environmental management is an extremely valuable characteristic of a result-based scheme such as our MIRBAP concept (see also 4.6 below). Although difficult to quantify, it is nevertheless potentially easy to eliminate the potential for engagement through insufficient care for farmers' perceptions of a system design.

**4.5 Dynamic efficiency**

An obvious downside of action-based payments is that they offer few incentives to innovate – since the payment is tied to a specific management action, it is rational for the participating farmer to stick to this action. Conversely, since the farmer is rewarded on an annually recurring basis for (measured or predicted) results, a result-based payment creates an incentive to consider ways of improving the effectiveness of environmentally beneficial management practices in the future, because the more effective a given practice or the lower the cost of implementing it, the higher the farmer's future profits. The effectiveness of practices can be influenced in two ways. Either through innovations that improve the practices or through technical changes in agriculture that indirectly affect the environmental performance of a management practice, where the effect could be negative. There is broad evidence that environmental taxes and payments in other sectors that are based on results, e.g. air-pollution taxes, are a catalyst for technical change that has substantially reduced the costs of environmental improvements, while action-based schemes have the least potential to promote dynamic efficiency (Requate, 2005). Furthermore, any result-based scheme has greater potential to engage farmers in environmental stewardship by offering a pecuniary reward for being knowledgeable of environmental impacts or being innovative. Farmers have over the centuries shown great creativity and innovation to improve the effectiveness of agriculture, because their livelihood has depended on it. Given that a MIRBAP scheme creates a link between environmental effectiveness and livelihoods, one can expect the scheme to promote the desire for learning (through, e.g., experimenting with the model) and harness their innovative capacity for resolving environmental problems.

Nevertheless, MIRBAP scores less well in terms of dynamic efficiency than conventional result-based schemes that rely on measurement. Farmers cannot benefit from innovation if the environmental benefits from such innovation cannot be modelled. Innovation is only theoretically possible if a farmer suggests something that can be modelled easily. However, in general it takes several years of research to develop a model to predict effects of practices with acceptable uncertainty. More realistically, farmers can instead be engaged in the model-building process and contribute before the scheme is launched. It has been shown that stakeholders who participate in model building for environmental decision-making contribute with information, novel ideas and solutions (Beierle, 2002), increased acceptance and uptake of modelled results (Wassen et al., 2011) and improved environmental outcomes (Brody, 2003). This can also help the modellers understand the context and management practices for which they model the outcomes, while also helping the policy makers and farmers understand what is feasible to model and under what uncertainty (Addison et al., 2013). Moreover, environmental management, as well as agricultural management are annually recurring processes. If farmers are provided with a mechanism to feed into model improvement, so that they can expect innovations to be included in the model over time, then the system would promote dynamic efficiency, though with a time lag.

Time lags are characteristic of the diffusion of technological developments in agriculture generally, hence the inability for immediate introduction of environmental innovations should not pose a significant disadvantage, if the long-term advantages can be perceived by farmers. A mechanism to involve farmers in further model development (see step 5 in Figure 1) could also promote farmer collaboration and the sharing of ideas, which is known to be important for innovation (Darnhofer et al., 2010; Mills et al., 2019). For example, if the introduction of a proposal for model improvement requires a minimum level of consensus among farmers, it should encourage dialogue amongst farmers and modellers through e.g. facilitation activities. Conversely, it is an open question how willing farmers would be to "play around" with the MIRBAP tool and to participate in its further development (e.g. adding new practices to the modelled portfolio). The challenge here is mainly how to best give opportunities to farmers and other stakeholders to provide input into further model development.

Relatedly, it is important to minimize reporting and data input required from farmers to feed the model – in some cases GIS layers with currently available data may be too imprecise for application in heterogeneous areas. Given the rise of precision farming technologies and digitalization of agriculture (Finger et al., 2019; Weersink et al., 2018), data may be provided by the farmers themselves. Then, standards would be needed to ensure valid model results.

### 4.6 Autonomy and non-pecuniary motives

It is a general advantage of result-based schemes that they are consistent with farmers' preference for making their own decisions about how to manage their land, including management for environmental objectives. Thus, result-based schemes not only give room for immaterial benefits of a feeling of agency and autonomy, but also allow for using relevant local knowledge (e.g. Riley, 2016; Stupak et al., 2019). In this respect, MIRBAP has both a relative advantage and a disadvantage as compared to measurement-based schemes.

First, the MIRBAP interface is not only useful in the context of the payment scheme itself – but also has the character of a decision support tool, though one focusing strongly on environmental effects of agricultural management practices (cf. Fales et al., 2016). In this sense, MIRBAP has the appeal of combining a policy instrument with a decision support tool that would support the farmers in their pursuit of both private and public goals. Furthermore, by linking management choices to their environmental effects, MIRBAP would facilitate the consideration of non-pecuniary motives by farmers, who may make decisions not only on the basis of profit maximization (Bartkowski and Bartke, 2018). This is a significant advantage as compared to action-based payment schemes, in which knowledge about the effects of practices by individual farmers is not usually available to them, so it is rational for the farmer to focus on pecuniary rewards. Having access to the model would enable the farmer to experiment and thereby learn from the predicted results of different actions. Normally, farmers are not provided with individual feedback on the environmental consequences of their management choices in an action-based scheme, which deadens engagement. Further, this step is crucial for building trust in the system, because it is more transparent than a government expert running the model and telling farmers what they can do (as we also point out in relation to cost effectiveness in section 4.4), instead of farmers telling the government what they are willing to do.

However, for reasons already discussed with respect to dynamic efficiency, MIRBAP is still more restrictive than a conventional result-based scheme because it only allows for management changes that are and can be modelled. Therefore, the autonomy of the participating farmers is not as high, though still higher than in the case of action-based schemes.

### 4.7 Multiple and long-term objectives

There is an increasing number of models that allow taking into account multiple environmental effects (e.g. ecosystem service bundles), thus supporting the analysis of trade-offs between different objectives. Using such multi-objective models to support a MIRBAP scheme (see also section 3.2) would allow regulators to provide more "holistic" incentives, focusing on the interactions between different environmental objectives rather than treating each of them separately (or some of them not at all, e.g. due to prohibitive costs of measurement). Furthermore, it would enable and support farmers in choosing management approaches that improve multiple ecosystem services simultaneously – while also making them aware of the trade-offs involved. Relatedly, the MIRBAP framework offers an option for targeting long-term processes, for which measurement would imply long time lags. Given the usual contract length of five years in the EU, no management practices are currently incentivized whose positive environmental impacts take longer. For instance, the effects of practices changing the structural development of soils to improve water infiltration and storage or the consequences of crop diversification as a substitute for pesticides on soil biology may be detectable only after longer periods. Modelling allows predicting the far-into-the-future effects of various practices and provide remuneration to farmers accordingly, thus also giving them incentives for longer-term investments in changing practices. This would further increase the range of applications and their societal relevance in agri-environmental policy. Of course, it would also require longer contract lengths to ensure that the changes are not reversed before taking effect.

## 5 Conclusions

In this paper, we introduced a novel conceptual idea for the design of agri-environmental payment schemes – model-informed result-based agri-environmental payments (MIRBAP). MIRBAP is a combination of design elements, but also of most advantages of conventional result-based payments and the payment certainty of action-based schemes. On the one hand, the prime advantages of the former would be retained – high environmental effectiveness (outcome certainty), cost effectiveness, dynamic efficiency and facilitation of farmers' autonomy. Given high quality models, it can be assumed that on average, the predicted results will be realized (with some random variation due to factors such as weather conditions). Importantly, the two main practical downsides of result-based schemes (vis-à-vis action-based), namely costs of measurement and payment uncertainty for farmers, can be resolved with the proposed modelling approach. In the first instance, the need to visit all fields and carry out expensive analysis is replaced by modelling results, with measurements restricted to a sample of fields for either continual model validation and improvement, or regulatory control. Furthermore, since the model would predict the environmental effects ex ante and the payments to the farmer would be based on these predictions, there would be no payment uncertainty for the farmer. In this sense, MIRBAP has the potential to reduce outcome uncertainty as compared with action-based schemes (from the regulator's perspective), while also reducing payment uncertainty as compared with conventional result-based schemes (from the farmer's perspective). Overall, the MIRBAP scheme would thus improve social welfare. The payment would be tied to environmental outcomes and thus would be cost-effective since society would only pay for what is actually obtained and farmers would seek the least costly measures to obtain the payments. The downside from society's perspective is that if actual results are lower, the real effectiveness of the scheme is reduced. This downside is minimized over time in our framework through the design element of continual model validation and development.

Two major improvements that go beyond conventional action-based or result-based schemes are the possibility to address multiple objectives and long-term objectives. As illustrated by referring to the modelling framework under development in the BonaRes project, MIRBAP would increase the policy relevance of more complex, multi-objective models, e.g. in the generally rather neglected context of soil functions. Furthermore, using models also allows to take on a longer-term perspective and base payments today on effects that are expected farther in the future.

Our paper offers the conceptual outline of MIRBAP. For it to become a viable option for policy, there is a need for further research. First, farmers' acceptance of and willingness to participate in a MIRBAP scheme should be studied – the experience reported by Fales et al. (2016) provides first tentative reasons for optimism. Second, the relevance of various models and modelling frameworks for MIRBAP should be tested in more detail – the framework has the largest potential where measurement is difficult or the achievement of the scheme's goal is highly uncertain for farmers. Third, there is a need for new ways of increasing model robustness and flexibility, so as to allow uptake of innovations (to spur dynamic efficiency). Fourth, interface solutions should be developed to maximize the usability and attractiveness of MIRBAP in practice. This should be informed by farmers' preferences. Above all, however, there is a need for a pilot study applying the MIRBAP principles in a real-world context.

## Acknowledgements

We would like to thank Sophie Binder for compiling Table A1. We also would like to thank Frank Wätzold for unwittingly initiating this research by bringing the authors together. BB, ML and UW acknowledge funding from the German Federal Ministry of Education and Research (BMBF) in the framework of the funding measure "Soil as a Sustainable Resource for the Bioeconomy – BonaRes", project "BonaRes (Module B): BonaRes Centre for Soil Research, subproject A" (grant 031B0511A). The research presented in this paper is a contribution to the strategic research area Biodiversity and Ecosystems in a Changing Climate, BECC, through the project BioInsure.

# Appendix

Table A1: Overview result-based schemes in Europe [source: own elaboration based on EC's inventory, http://ec.europa.eu/environment/nature/rbaps/fiche/map_en.htm]

| Scheme | Country/region | Indicator |
|---|---|---|
| Golden Eagle conservation scheme | Finland | Eagle nests & chicks |
| Conservation Performance payments | Sweden | Number of lynx and wolverine offspring |
| Burren Farming for Conservation Programme | Ireland | Fields scored (0–10) based on: grazings levels, feeding systems, scrub and weed encroachment, condition of water sources, site integrity |
| Farm Conservation Scheme | England, UK | Indicator plant species (separated into five different groups) |
| Grassland Bird Protection Payments (Gemeinschaftlicher Wiesenschutz) | Schleswig-Holstein, Germany | Presence of breeding birds, number of clutches per hectare |
| Harrier nest protection in arable fields | Nordrhein-Westfalen, Germany | Presence of nests of Montague's Harrier or Marsh Harrier |
| Maintenance of species rich grassland | Germany, various federal states | Minimum number of four indicator species from a regional list present (or more for higher payments) |
| Maintenance of traditional orchards | Germany, various federal states | Payment per tree (nut or fruit) of a certain minimum size to a max density of 100 trees per ha |
| Species rich grassland | Rheinland-Pfalz, Germany | Presence of a minimum of 4–8 indicator species from catalogue in each third of a transect across the parcel |
| On-farm conservation of rare | EU | Number of heads of breeding female or |

| and endangered local animal breeds | | male animals and offspring (with breeding recorded) |
|---|---|---|
| Species rich grassland scheme (MEKA programme B4) | Baden-Württemberg, Germany | Presence of a minimum of 4 indicator species (28 key flower species or genera) from catalogue in each third of a transect across the parcel |
| Results-based nature conservation plan (Ergebnsiorientierter Naturschutzplan) | Austria | Individually set results and control criteria for each parcel (High Nature Value farmland) |
| Species rich vineyards (Rebflächen mit natürlicher Arten- und Strukturvielfalt) | Switzerland | Individual scores: indicator species and structural criteria (stone walls, individual trees, wildflower areas) |
| Preservation and enhancement of species rich grassland (Öko-Qualitätsverordnung ÖQV – Ecological Compensation Areas) | Switzerland | Presence of a minimum of 6 indicator species from catalogue in a 6 metre diameter circle |
| Species rich grassland programme (Flowering Meadows Scheme) | France | Presence of (a minimum of) 4 plant species from a list |
| Pastoral management plan (Gestion pastorale) | France | Change in vegetation state (assessed against photographic reference information), management actions to be undertaken by farmers defined at each site |
| RAPCA (Red de Áreas Pasto-Cortafuegos de Andalucía (managing firebreaks)) | Andalucía, Spain | Evaluation of overall vegetation structure, visual assessment of how much of individual shrubs have been consumed, overall consumption of herbaceous layer |
| Per Clutch Trials | Netherlands | Number of clutches |
| Meadow bird agreement with agri-environment cooperatives | Netherlands | Number of bird nests |
| Meadow bird agreements (2000-2006) | Netherlands | Number of breeding pairs per ha |
| Species rich grassland and arable botanical management agreements (2000-2006) | Netherlands | Number of domestic plant species according to agreement (10, 15, 20 different grassland species per 25 sqm) |